\documentclass[aps,prl,groupedaddress]{revtex4-1}
\usepackage{graphicx}

\begin{document}


\title{Single shot ultrafast laser processing of high-aspect ratio nanochannels using elliptical Bessel beams}


\author{R. Meyer} 
\author{M. Jacquot} 
\author{R. Giust}
\author{J. Safioui}
\author{L. Rapp} 
\author{L. Furfaro}
\author{P.-A. Lacourt}
\author{J.M. Dudley}
\author{F. Courvoisier}
\email[]{francois.courvoisier@femto-st.fr}
\affiliation{FEMTO-ST,Universit\'e de Bourgogne-Franche-Comt\'e UMR-6174, 25030 Besancon, France}


\date{\today}

\begin{abstract}
Ultrafast lasers have revolutionized material processing, opening a wealth of new applications in many areas of science. A recent technology that allows the cleaving of transparent materials via non-ablative processes is based on focusing and translating a high-intensity laser beam within a material to induce a well-defined internal stress plane. This then enables material separation without debris generation. Here, we use a non-diffracting beam engineered to have a transverse elliptical spatial profile to generate high aspect ratio elliptical channels in glass of dimension 350 nm × 710 nm, and subsequent cleaved surface uniformity at the sub-micron level. 
\end{abstract}

\maketitle


The high peak power of ultrafast laser pulses has revolutionized material processing, opening a wealth of new applications in microelectronics and for the development of technologies such as touch-screens and solar panels \cite{Gattass2008}. Although these mass markets for fabrication require material processing at high resolution and high speed, mechanical dicing or conventional laser ablation only poorly address this need. For example, mechanical techniques such as diamond saw dicing, water-jet dicing or scribe-and-cleave technique are slow and lead to excessive material waste, whilst conventional laser ablation generates deleterious surface debris \cite{Amoruso1999}. In this context, a new processing technology of transparent materials known as “stealth” machining is currently showing great promise because, rather than using ultrashort pulses for direct material ablation, infrared pulses are used to create an internal damage plane allowing cleaving without debris generation \cite{Kumagai2007}. In previous work, the damage plane was generated through laser-induced defects of $\sim$5-10 µm that induce internal cracks whose orientation could be controlled with elliptical beam shaping \cite{Dudutis2016, Hendricks2017}.

To increase the surface quality beyond the micron scale, stealth nanomachining has also recently been demonstrated using a series of aligned nanochannels processed by ultrafast lasers. However, whilst the promise of stealth nanomachining has been demonstrated using extended plasma filaments with symmetric transverse profiles \cite{Ahmed2008,Ahmed2013,Bhuyan2015,Mishchik2016} the cleaved surfaces did not precisely follow the linear array of machined channels, resulting in unwanted damage on either sides of the separated surfaces. In this paper, we solve this problem and report a new approach to stealth nanomachining using a novel class of “non-diffracting” – or propagation-invariant – beams that are engineered to have a transverse elliptical spatial profile. We show that the nonlinear filamentation of these beams generates high aspect ratio nanochannels in glass with an elliptical cross section, and this nano-scale ellipticity leads to well-defined stress planes which facilitates material separation and yields cleaved surfaces with record sub-micron uniformity.

\begin{figure}
 \includegraphics{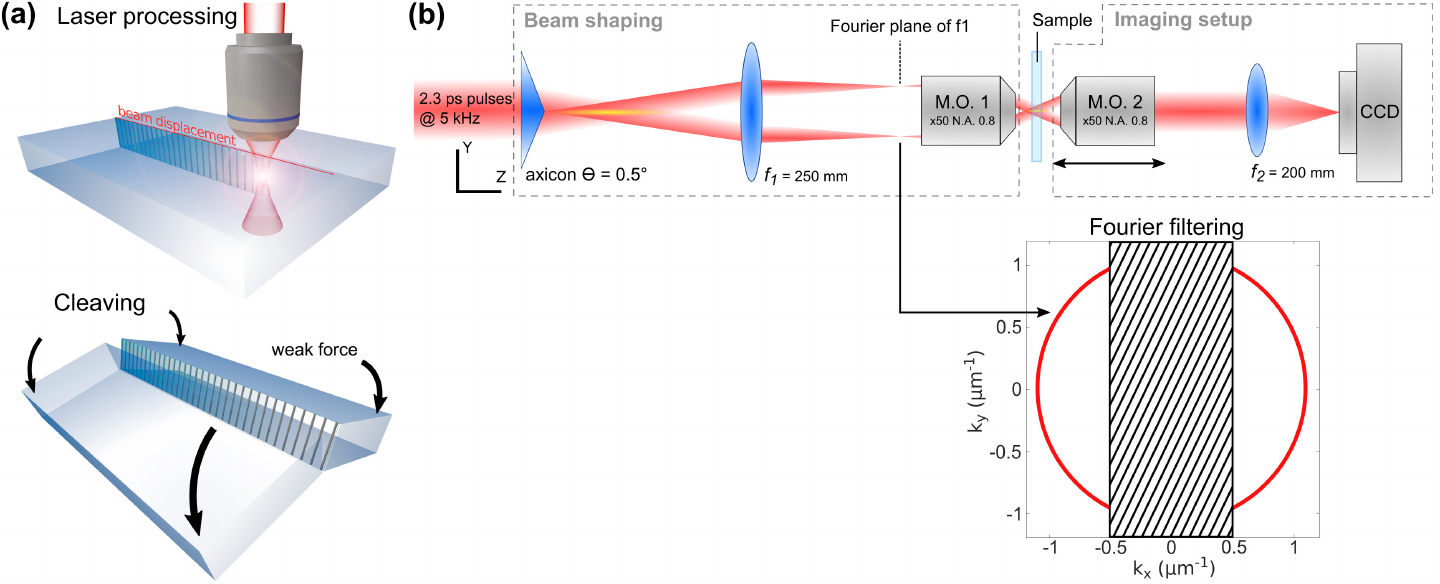}
	\caption{(a) Illustration of the concept of stealth nanomachining. (b) Experimental setup for beam shaping and beam imaging. The primary Bessel beam generated from the axicon is expanded and spatially filtered in the Fourier plane before focusing to sub-micron dimensions with a microscope objective (MO1, × 50, NA 0.8). The shaded area corresponds to the region where we block light to remove a specific range of spatial frequencies. The 3D beam intensity distribution is reconstructed after scanning the beam with a second microscope objective (MO2).}
	\label{fig:1}
\end{figure}
The physics of ultrafast laser material processing is based on the fact that the high peak power and picosecond or femtosecond duration of an incident pulse results in strongly-localized energy deposition at a confined site within a transparent material \cite{Gattass2008}. This energy deposition can be sufficiently high to induce a strong internal stress field \cite{Champion2013} and associated internal fissures and voids \cite{Ahmed2008,Sakakura2013,Juodkazis2006,Rapp2016,Gotte2016}. A specific regime of energy deposition occurs when the pulse propagation is highly nonlinear and generates spatial filaments that generate a longitudinally extended region of damage within the depth of the transparent dielectric. The principle of stealth nanomachining is to combine the generation of such extended damage regions with high-speed scanning to induce a weakened plane within the material. When followed by the application of a weak applied mechanical stress, the material cleaves naturally along this plane resulting in material separation without material loss or deleterious debris \cite{Kumagai2007,Hendricks2017,Ahmed2013,Bhuyan2015,Mishchik2016}, in contrast with a conventional ablative process.

Although nonlinear filamentation can occur with any focused ultrashort pulse, such spontaneous filamentation is highly sensitive to variation of input conditions \cite{Couairon2007}. As a result, machined structures arising from spontaneous filamentation are difficult to predict \cite{Sudrie2002}. In this context, however, the use of non-diffracting Bessel beams presents significant advantages. Such Bessel beams are formed by the cylindrically symmetric interference of plane waves, which cross the optical axis at a given conical angle, forming a high-intensity central lobe. surrounded by circular lobes of reduced intensity \cite{McGloin2005}. Significantly, the diameter of the central lobe is invariant along the propagation direction and has been shown to induce a stable (stationary) regime of nonlinear filamentation in transparent dielectrics \cite{Polesana2008}, with homogeneous induced damage along all regions of the beam where the intensity exceeds the threshold for void formation \cite{Rapp2016,Bhuyan2010,Courvoisier2016}. Although the use of picosecond Bessel beams with symmetric transverse beam profiles for such material separation has been previously reported in \cite{Bhuyan2015}, high resolution Scanning Electron Microscopy (SEM) reveals that the cleaved surface does not precisely follow the line through the laser-drilled nano-channels. This is an important drawback because the channels left uncleaved act as weakening sites which increase the brittleness of the material after processing. To this end, we develop here a novel elliptical Bessel beam, where the transverse intensity distribution in the main lobe has an elliptical shape and which solves this problem.

\begin{figure}[htbp]
\includegraphics[width=0.7\linewidth]{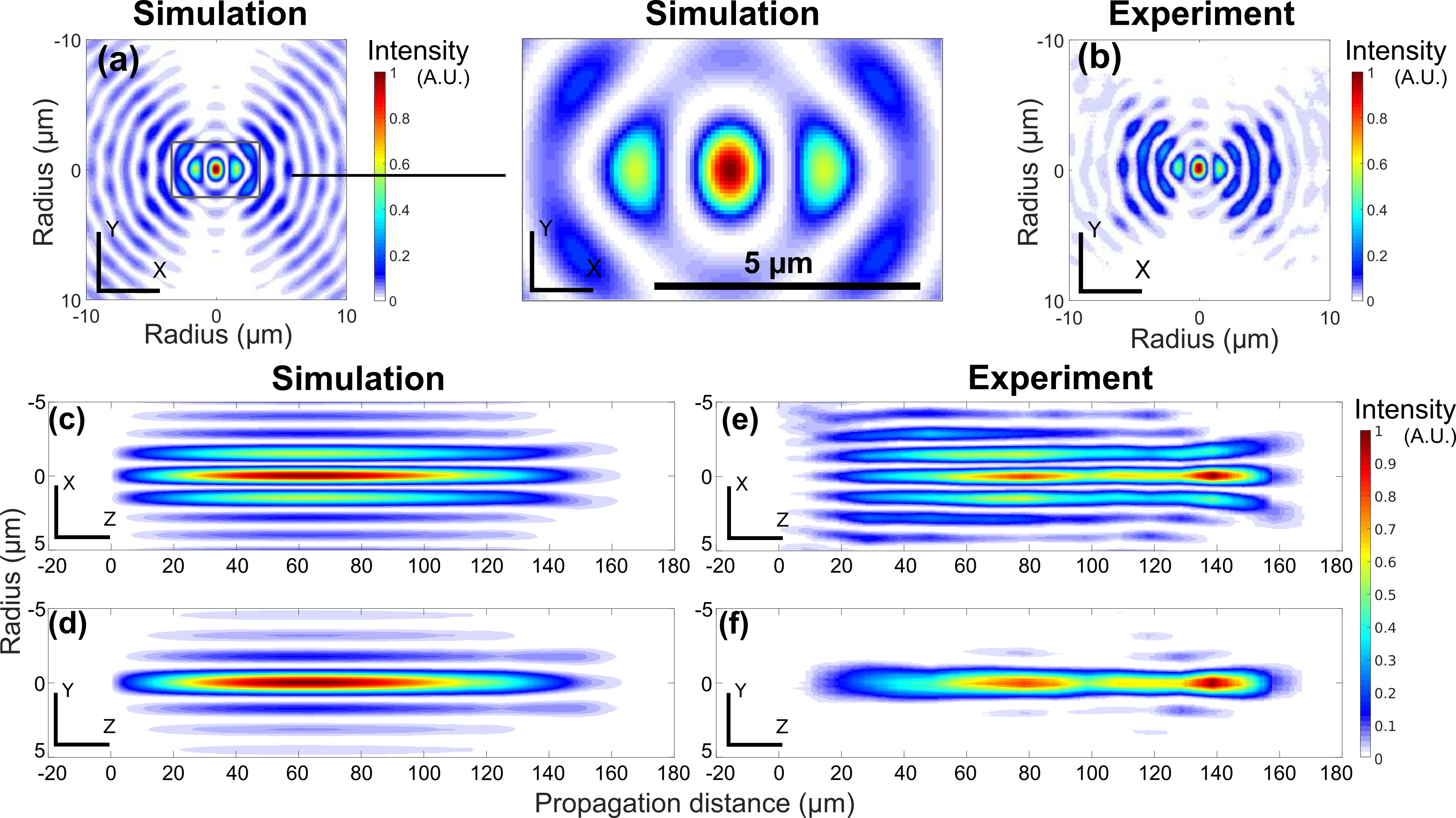}
\caption{Intensity distribution of the elliptical Bessel beam. (a) Numerical simulation of the beam cross section at propagation distance z=110 µm from the focal plane of MO1. The inset shows a magnified view of the central part of the beam, featuring its elliptical central lobes. (b) Corresponding experimental profile. (c) to (f) Longitudinal intensity distributions in air in the XZ and YZ planes. (c, d) are numerical simulations and (e, f) the corresponding experimental results.}
	\label{fig:2}
\end{figure}

The principle of our experiment is shown in the schematic illustration in Fig. \ref{fig:1}(a). A suitably shaped elliptical Bessel beam is used to generate a linear array of nanochannels inside a glass sample. This is achieved in our case using single-shot illumination of circularly-polarized 5 kHz amplified Ti:Sapphire laser pulses at 800 nm to generate each nanochannel while the sample is being linearly translated in the beam focus. Following this laser processing step, a weak applied force is used to cleave the sample along the induced stress plane. Figure \ref{fig:1}(b) shows more details of the experimental setup.

The grating position in the compressor of our Chirped Pulse Amplifier laser allowed us to vary the pulse duration in the range 120 fs-3 ps.  In a preliminary study with circularly symmetric Bessel beams, we determined that a pulse duration of 2.3 ps associated to sample translation speed of 25 mm/s was the best configuration for stealth nanomachining. The few picoseconds regime has indeed previously been shown to be more suitable for stealth nanomachining [8]. Circular polarization was generated with a quarter waveplate just after the laser source and the polarization is preserved through the rest of the setup \cite{Zhang2008}. The primary (circularly symmetric) Bessel beam is generated from a 0.5° axicon and without any additional beam shaping, demagnification leads to a circularly symmetric beam characterized by a cone angle of 16.5° and a central spot size of $\sim$1.0 µm FWHM extending over a longitudinal distance of 170 µm (in air, without any sample present). Ellipticity is introduced in the transverse profile by Fourier filtering of the primary Bessel beam as illustrated in the exploded view shown in Fig. \ref{fig:1}(b). Specifically, the removal of spectral components along a line in the y-direction yields a non-circularly symmetric intensity distribution with three strongly elliptical lobes. The effect of this filtering is easily calculated numerically and is shown in Fig. \ref{fig:2}(a) at a distance of z = 110 µm from the focal plane of the microscope objective MO1 (distance where the sample would be positioned). A zoom of the central region is also shown to highlight the strong ellipticity of both the main intensity lobe and the side lobes. These calculated results are confirmed in the experimental characterization shown in Fig.  \ref{fig:2}(b), obtained by translating a second microscope objective to reconstruct the 3D field plane-by-plane. Note that the filtering induces energy loss of $\sim$30\% but that all energy values given below are those measured at the sample site.

It is also possible to readily calculate (using the standard scalar beam propagation method for Fourier optics \cite{Goodman1996} the expected longitudinal evolution of the elliptical Bessel beam and this is shown in Fig \ref{fig:2}. (c,d) for the XZ and YZ planes respectively (in air). Corresponding experimental results are shown in Fig.  \ref{fig:2}(e,f). We clearly see the expected non-diffracting behavior of the shaped Bessel beam over a longitudinal extent of more than 100 µm. The experimental results do show some longitudinal non-uniformity but this has negligible effect on the generation of machined nanochannels provided the intensity exceeds the ionization threshold at all points where the sample is positioned.

\begin{figure}[htbp]
\includegraphics[width=0.9\linewidth]{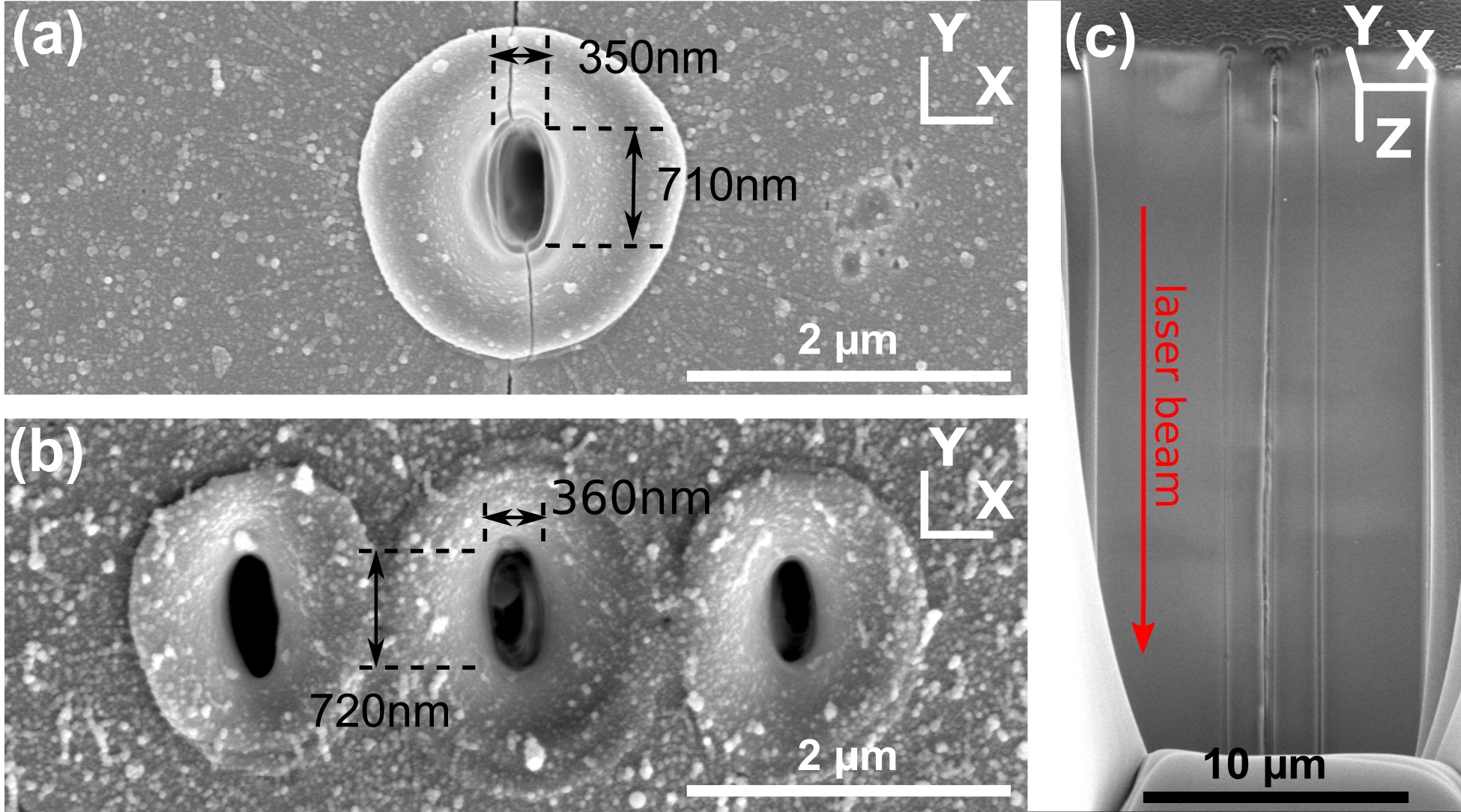}
\caption{SEM imaging of laser-induced channel using single shot illumination with an elliptical Bessel beam: (a) Image of top surface damage produced by illumination with a pulse energy of 12 µJ. A single elliptic nanochannel is visible at the center of round-shaped ejected material, placed in between two lateral index modifications (not visible under SEM) produced by the secondary lobes. (b) Image of top surface damage produced by illumination with a pulse energy of 18 µJ, showing the 3 nanochannels produced by the 3 main hot spots of the elliptical beam. (c) View of the longitudinal section of the channels shown in (b) after material removal by FIB milling over a depth of 30 µm. The laser beam is coming from top, sample is observed with a 52° tilt. The 3 parallel channels corresponding to the three main hot spots are apparent. }
	\label{fig:3}
\end{figure}

Results showing the use of this setup for generating nanochannels in glass are shown in Fig. \ref{fig:3}. Here, a 150 µm thick glass sample (Schott D263M borosilicate) is placed at the shaped Bessel beam focus, and through control of the incident pulse energy, we are able to generate either (a) one single channel (12 µJ energy when only the central lobe exceeds the ionization threshold) or (b) three channels simultaneously (18 µJ energy when all three central lobes exceed the void formation threshold). The results in Fig. \ref{fig:3}(a) and (b) show the surface features on the top side correspond to these two pulse energies, and the cross section in Fig. \ref{fig:3}(c) shows the longitudinal structure for the case of three channels after material using focused ion beam milling (FIB). Even with three lobes with intensities $~10^{13}-10^{14} W/cm^{2}$, the propagation shows stable beam propagation through the full thickness of the sample. This shows the elliptical Bessel beam preserves the main properties of propagation invariance of regular Bessel beams even in the nonlinear ionization regime \cite{Polesana2008,Bhuyan2010}. A careful FIB milling procedure allowed us to minimize the damage produced to the laser-produced channels \cite{Rapp2016}. High resolution FIB images show a variation in diameter of about $\sim$50 \% along the length of the channel, understood from inhomogeneity of the on-axis beam intensity, as can be seen in Fig \ref{fig:2}(f).

\begin{figure}[htbp]
\includegraphics[width=0.78\linewidth]{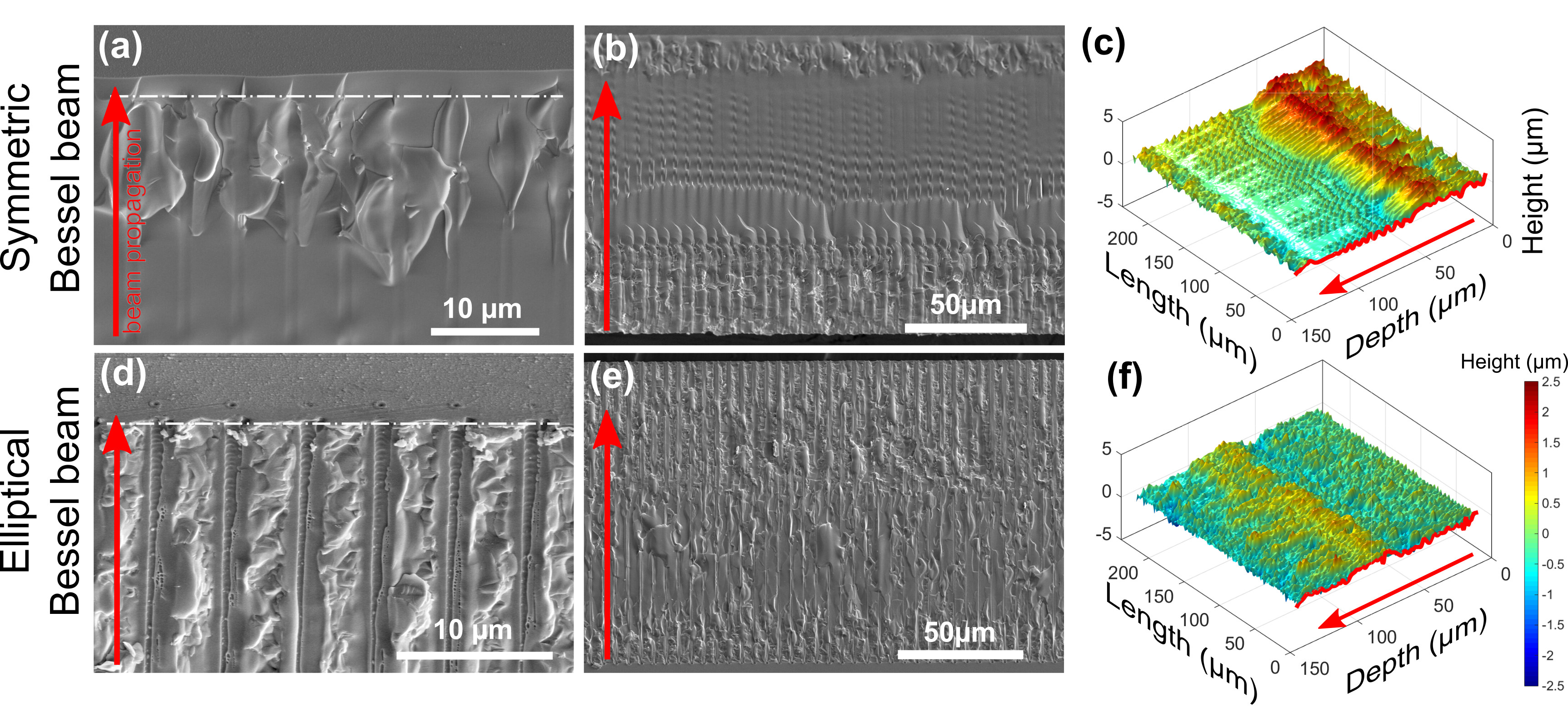}
\caption{ Cleaved edge profiles of samples laser-processed with circularly symmetric (top line) and elliptical (bottom line) Bessel beams with pulse energy of 12 µJ. Red arrows show the beam propagation direction. (a,d) SEM 52° tilted views at sample exit face; white dot-dash line shows the expected fracture line. (b,e) SEM 52° tilted full views of the samples cleaved edges. In (e), nano-channels are open from top surface to exit surface. (c,f) Topography measurement of the edge surface with optical profilometry featuring flatter surface profiles for the elliptical Bessel beam processing (f).}
	\label{fig:4}
	\end{figure}

The nanochannel ellipticity is an important feature for applications to stealth processing where it is the induced material stress that enables the definition of a well-defined cleaving plane \cite{Collins2015}. In particular, an elliptical hole in a solid structure enhances the stress contrast between the cleaving plane and the sides of the perforation, which can be approximately quantified through the stress concentration factor given by $\kappa = 1 + 2\epsilon$ for a hole of eccentricity $\epsilon$ \cite{Young2002}. This yields $\kappa = 5$ for elliptical hole with $\kappa = 2$ compared to $\kappa = 3$ for a circular hole, suggesting a 66\% stress enhancement. This stress enhancement is also much more localized near the tips of the elliptical profile which would be expected to lead to a significantly more well-defined stress plane.

We can also compare the properties of the machined samples when using circular and elliptical beams. After single pass illumination along the y-axis at a constant speed of 25 mm/s, the samples were mechanically bent using a 3 point-flexure setup until they cleaved \cite{Nattermann1998}. Samples processed with an elliptical Bessel beam need typically 50 \% less deflection for cleaving than those processed with a circular Bessel beam. We compare in Fig. \ref{fig:4} the samples under SEM microscopy. The left column shows the 3D view and side view of the sample processed after circular Bessel beam illumination. It is apparent that the cleavage occurred to one side of the line of channels (dashed white line) by a few micrometers. A full characterization over the samples reveals that the cleavage occurs randomly on either side of the line of the nanochannels. This is quantitatively confirmed by optical profilometry measurements as shown in Fig. \ref{fig:4}(c). In contrast, the elliptical Bessel beam processing produced samples where cleaving occurred precisely along the line defined by the vertices of the ellipses due to the stress enhancement. The position of the cleaved surface follows the line to within less than 1 µm all along the length and depth of the 10/10 samples characterized. We note that no difference was observed if the pressure was applied on top or exit side of the samples for circularly symmetric and elliptical Bessel beams.  We also stress that the results in Fig. \ref{fig:4} with symmetric and elliptical beams are obtained at the same energy, and the apparent reduced damage seen with symmetric beams is because cleaving occurs slightly off-centre in the channels. After averaging over the cleaved surface area at 12 µJ, surface roughness (root mean square) is lower for results using an elliptical Bessel beam ($\sim$0.5 µm) than a circular Gaussian beam ($\sim$0.6 µm).

In conclusion, we have reported how the use of novel transverse beam shaping of non-diffracting beams can yield a propagation invariant regime at ablation-level intensities that generates elliptical nanochannels in glass after single shot illumination. We have demonstrated that this property can be used to guide material cleaving with sub-micron precision. We also note that while nanochannels can be produced by femtosecond and picosecond illumination, only the longest pulse durations produce cleavable damages. This indicates that the thermal modification around the nanochannels plays a role in the fracture propagation. Additional work is needed to fully uncover the influence on cleaving of the nanochannel shape and density distribution around the channel. We anticipate that these results will foster further novel research in the nonlinear propagation of intense pulses with structured profiles, and new applications in laser materials processing.

\section*{Funding}

European Union Seventh Framework Programme [ICT 2013.3.2 Photonics], grant agreement n°619177, TiSaTD; European Research Council (ERC), grant agreement n° 682032-PULSAR; Labex ACTION program, contract ANR-11-LABX-0001-01 

\section*{Acknowledgements}
This work was partly supported by the French RENATECH network
\bibliography{StealthDicingBib}

\begin{thebibliography}{25}%
\makeatletter
\providecommand \@ifxundefined [1]{%
 \@ifx{#1\undefined}
}%
\providecommand \@ifnum [1]{%
 \ifnum #1\expandafter \@firstoftwo
 \else \expandafter \@secondoftwo
 \fi
}%
\providecommand \@ifx [1]{%
 \ifx #1\expandafter \@firstoftwo
 \else \expandafter \@secondoftwo
 \fi
}%
\providecommand \natexlab [1]{#1}%
\providecommand \enquote  [1]{``#1''}%
\providecommand \bibnamefont  [1]{#1}%
\providecommand \bibfnamefont [1]{#1}%
\providecommand \citenamefont [1]{#1}%
\providecommand \href@noop [0]{\@secondoftwo}%
\providecommand \href [0]{\begingroup \@sanitize@url \@href}%
\providecommand \@href[1]{\@@startlink{#1}\@@href}%
\providecommand \@@href[1]{\endgroup#1\@@endlink}%
\providecommand \@sanitize@url [0]{\catcode `\\12\catcode `\$12\catcode
  `\&12\catcode `\#12\catcode `\^12\catcode `\_12\catcode `\%12\relax}%
\providecommand \@@startlink[1]{}%
\providecommand \@@endlink[0]{}%
\providecommand \url  [0]{\begingroup\@sanitize@url \@url }%
\providecommand \@url [1]{\endgroup\@href {#1}{\urlprefix }}%
\providecommand \urlprefix  [0]{URL }%
\providecommand \Eprint [0]{\href }%
\providecommand \doibase [0]{http://dx.doi.org/}%
\providecommand \selectlanguage [0]{\@gobble}%
\providecommand \bibinfo  [0]{\@secondoftwo}%
\providecommand \bibfield  [0]{\@secondoftwo}%
\providecommand \translation [1]{[#1]}%
\providecommand \BibitemOpen [0]{}%
\providecommand \bibitemStop [0]{}%
\providecommand \bibitemNoStop [0]{.\EOS\space}%
\providecommand \EOS [0]{\spacefactor3000\relax}%
\providecommand \BibitemShut  [1]{\csname bibitem#1\endcsname}%
\let\auto@bib@innerbib\@empty
\bibitem [{\citenamefont {Gattass}\ and\ \citenamefont
  {Mazur}(2008)}]{Gattass2008}%
  \BibitemOpen
  \bibfield  {author} {\bibinfo {author} {\bibfnamefont {R.~R.}\ \bibnamefont
  {Gattass}}\ and\ \bibinfo {author} {\bibfnamefont {E.}~\bibnamefont
  {Mazur}},\ }\href {\doibase 10.1038/nphoton.2008.47} {\bibfield  {journal}
  {\bibinfo  {journal} {Nat. Photon.}\ }\textbf {\bibinfo {volume} {2}},\
  \bibinfo {pages} {219} (\bibinfo {year} {2008})}\BibitemShut {NoStop}%
\bibitem [{\citenamefont {Amoruso}\ \emph {et~al.}(1999)\citenamefont
  {Amoruso}, \citenamefont {Bruzzese}, \citenamefont {Spinelli},\ and\
  \citenamefont {Velotta}}]{Amoruso1999}%
  \BibitemOpen
  \bibfield  {author} {\bibinfo {author} {\bibfnamefont {S.}~\bibnamefont
  {Amoruso}}, \bibinfo {author} {\bibfnamefont {R.}~\bibnamefont {Bruzzese}},
  \bibinfo {author} {\bibfnamefont {N.}~\bibnamefont {Spinelli}}, \ and\
  \bibinfo {author} {\bibfnamefont {R.}~\bibnamefont {Velotta}},\ }\href@noop
  {} {\bibfield  {journal} {\bibinfo  {journal} {Journal of Physics B: Atomic,
  Molecular and Optical Physics}\ }\textbf {\bibinfo {volume} {32}},\ \bibinfo
  {pages} {R131} (\bibinfo {year} {1999})}\BibitemShut {NoStop}%
\bibitem [{\citenamefont {Kumagai}\ \emph {et~al.}(2007)\citenamefont
  {Kumagai}, \citenamefont {Uchiyama}, \citenamefont {Ohmura}, \citenamefont
  {Sugiura}, \citenamefont {Atsumi},\ and\ \citenamefont
  {Fukumitsu}}]{Kumagai2007}%
  \BibitemOpen
  \bibfield  {author} {\bibinfo {author} {\bibfnamefont {M.}~\bibnamefont
  {Kumagai}}, \bibinfo {author} {\bibfnamefont {N.}~\bibnamefont {Uchiyama}},
  \bibinfo {author} {\bibfnamefont {E.}~\bibnamefont {Ohmura}}, \bibinfo
  {author} {\bibfnamefont {R.}~\bibnamefont {Sugiura}}, \bibinfo {author}
  {\bibfnamefont {K.}~\bibnamefont {Atsumi}}, \ and\ \bibinfo {author}
  {\bibfnamefont {K.}~\bibnamefont {Fukumitsu}},\ }\href@noop {} {\bibfield
  {journal} {\bibinfo  {journal} {IEEE Transactions on Semiconductor
  Manufacturing}\ }\textbf {\bibinfo {volume} {20}},\ \bibinfo {pages} {259}
  (\bibinfo {year} {2007})}\BibitemShut {NoStop}%
\bibitem [{\citenamefont {Dudutis}\ \emph {et~al.}(2016)\citenamefont
  {Dudutis}, \citenamefont {Ge{\v{c}}ys},\ and\ \citenamefont
  {Ra{\v{c}}iukaitis}}]{Dudutis2016}%
  \BibitemOpen
  \bibfield  {author} {\bibinfo {author} {\bibfnamefont {J.}~\bibnamefont
  {Dudutis}}, \bibinfo {author} {\bibfnamefont {P.}~\bibnamefont
  {Ge{\v{c}}ys}}, \ and\ \bibinfo {author} {\bibfnamefont {G.}~\bibnamefont
  {Ra{\v{c}}iukaitis}},\ }\href@noop {} {\bibfield  {journal} {\bibinfo
  {journal} {Optics Express}\ }\textbf {\bibinfo {volume} {24}},\ \bibinfo
  {pages} {28433} (\bibinfo {year} {2016})}\BibitemShut {NoStop}%
\bibitem [{\citenamefont {Hendricks}\ \emph {et~al.}(2017)\citenamefont
  {Hendricks}, \citenamefont {Bernard},\ and\ \citenamefont
  {Matylitsky}}]{Hendricks2017}%
  \BibitemOpen
  \bibfield  {author} {\bibinfo {author} {\bibfnamefont {F.}~\bibnamefont
  {Hendricks}}, \bibinfo {author} {\bibfnamefont {B.}~\bibnamefont {Bernard}},
  \ and\ \bibinfo {author} {\bibfnamefont {V.}~\bibnamefont {Matylitsky}}\
  }(\bibinfo {year} {2017})\ pp.\ \bibinfo {pages} {10094 --
  1009417}\BibitemShut {NoStop}%
\bibitem [{\citenamefont {Ahmed}\ \emph {et~al.}(2008)\citenamefont {Ahmed},
  \citenamefont {Lee}, \citenamefont {Sekita}, \citenamefont {Sumiyoshi},\ and\
  \citenamefont {Kamata}}]{Ahmed2008}%
  \BibitemOpen
  \bibfield  {author} {\bibinfo {author} {\bibfnamefont {F.}~\bibnamefont
  {Ahmed}}, \bibinfo {author} {\bibfnamefont {M.~S.}\ \bibnamefont {Lee}},
  \bibinfo {author} {\bibfnamefont {H.}~\bibnamefont {Sekita}}, \bibinfo
  {author} {\bibfnamefont {T.}~\bibnamefont {Sumiyoshi}}, \ and\ \bibinfo
  {author} {\bibfnamefont {M.}~\bibnamefont {Kamata}},\ }\href {\doibase
  10.1007/s00339-008-4672-2} {\bibfield  {journal} {\bibinfo  {journal} {Appl.
  Phys. A}\ }\textbf {\bibinfo {volume} {93}},\ \bibinfo {pages} {189}
  (\bibinfo {year} {2008})}\BibitemShut {NoStop}%
\bibitem [{\citenamefont {Ahmed}\ \emph {et~al.}(2013)\citenamefont {Ahmed},
  \citenamefont {Ahsan}, \citenamefont {Lee},\ and\ \citenamefont
  {Jun}}]{Ahmed2013}%
  \BibitemOpen
  \bibfield  {author} {\bibinfo {author} {\bibfnamefont {F.}~\bibnamefont
  {Ahmed}}, \bibinfo {author} {\bibfnamefont {M.~S.}\ \bibnamefont {Ahsan}},
  \bibinfo {author} {\bibfnamefont {M.~S.}\ \bibnamefont {Lee}}, \ and\
  \bibinfo {author} {\bibfnamefont {M.~B.~G.}\ \bibnamefont {Jun}},\ }\href
  {\doibase 10.1007/s00339-013-7705-4} {\bibfield  {journal} {\bibinfo
  {journal} {Appl. Phys. A}\ }\textbf {\bibinfo {volume} {114}},\ \bibinfo
  {pages} {1161} (\bibinfo {year} {2013})}\BibitemShut {NoStop}%
\bibitem [{\citenamefont {Bhuyan}\ \emph {et~al.}(2015)\citenamefont {Bhuyan},
  \citenamefont {Jedrkiewicz}, \citenamefont {Sabonis}, \citenamefont
  {Mikutis}, \citenamefont {Recchia}, \citenamefont {Aprea}, \citenamefont
  {Bollani},\ and\ \citenamefont {Trapani}}]{Bhuyan2015}%
  \BibitemOpen
  \bibfield  {author} {\bibinfo {author} {\bibfnamefont {M.~K.}\ \bibnamefont
  {Bhuyan}}, \bibinfo {author} {\bibfnamefont {O.}~\bibnamefont {Jedrkiewicz}},
  \bibinfo {author} {\bibfnamefont {V.}~\bibnamefont {Sabonis}}, \bibinfo
  {author} {\bibfnamefont {M.}~\bibnamefont {Mikutis}}, \bibinfo {author}
  {\bibfnamefont {S.}~\bibnamefont {Recchia}}, \bibinfo {author} {\bibfnamefont
  {A.}~\bibnamefont {Aprea}}, \bibinfo {author} {\bibfnamefont
  {M.}~\bibnamefont {Bollani}}, \ and\ \bibinfo {author} {\bibfnamefont
  {P.~D.}\ \bibnamefont {Trapani}},\ }\href {\doibase
  10.1007/s00339-015-9289-7} {\bibfield  {journal} {\bibinfo  {journal} {Appl.
  Phys. A}\ }\textbf {\bibinfo {volume} {120}},\ \bibinfo {pages} {443}
  (\bibinfo {year} {2015})}\BibitemShut {NoStop}%
\bibitem [{\citenamefont {Mishchik}\ \emph {et~al.}(2016)\citenamefont
  {Mishchik}, \citenamefont {Leger}, \citenamefont {Caulier}, \citenamefont
  {Skupin}, \citenamefont {Chimier}, \citenamefont {H\"{o}nninger},
  \citenamefont {Kling}, \citenamefont {Duchateau},\ and\ \citenamefont
  {Lopez}}]{Mishchik2016}%
  \BibitemOpen
  \bibfield  {author} {\bibinfo {author} {\bibfnamefont {K.}~\bibnamefont
  {Mishchik}}, \bibinfo {author} {\bibfnamefont {C.~J.}\ \bibnamefont {Leger}},
  \bibinfo {author} {\bibfnamefont {O.~D.}\ \bibnamefont {Caulier}}, \bibinfo
  {author} {\bibfnamefont {S.}~\bibnamefont {Skupin}}, \bibinfo {author}
  {\bibfnamefont {B.}~\bibnamefont {Chimier}}, \bibinfo {author} {\bibfnamefont
  {C.}~\bibnamefont {H\"{o}nninger}}, \bibinfo {author} {\bibfnamefont
  {R.}~\bibnamefont {Kling}}, \bibinfo {author} {\bibfnamefont
  {G.}~\bibnamefont {Duchateau}}, \ and\ \bibinfo {author} {\bibfnamefont
  {J.}~\bibnamefont {Lopez}},\ }\href {\doibase 10.2961/jlmn.2016.01.0012}
  {\bibfield  {journal} {\bibinfo  {journal} {J. Laser Micro/Nanoeng.}\
  }\textbf {\bibinfo {volume} {11}},\ \bibinfo {pages} {66} (\bibinfo {year}
  {2016})}\BibitemShut {NoStop}%
\bibitem [{\citenamefont {Champion}\ \emph {et~al.}(2013)\citenamefont
  {Champion}, \citenamefont {Beresna}, \citenamefont {Kazansky},\ and\
  \citenamefont {Bellouard}}]{Champion2013}%
  \BibitemOpen
  \bibfield  {author} {\bibinfo {author} {\bibfnamefont {A.}~\bibnamefont
  {Champion}}, \bibinfo {author} {\bibfnamefont {M.}~\bibnamefont {Beresna}},
  \bibinfo {author} {\bibfnamefont {P.}~\bibnamefont {Kazansky}}, \ and\
  \bibinfo {author} {\bibfnamefont {Y.}~\bibnamefont {Bellouard}},\ }\href
  {\doibase 10.1364/OE.21.024942} {\bibfield  {journal} {\bibinfo  {journal}
  {Opt. Express}\ }\textbf {\bibinfo {volume} {21}},\ \bibinfo {pages} {24942}
  (\bibinfo {year} {2013})}\BibitemShut {NoStop}%
\bibitem [{\citenamefont {Sakakura}\ \emph {et~al.}(2013)\citenamefont
  {Sakakura}, \citenamefont {Ishiguro}, \citenamefont {Fukuda}, \citenamefont
  {Shimotsuma},\ and\ \citenamefont {Miura}}]{Sakakura2013}%
  \BibitemOpen
  \bibfield  {author} {\bibinfo {author} {\bibfnamefont {M.}~\bibnamefont
  {Sakakura}}, \bibinfo {author} {\bibfnamefont {Y.}~\bibnamefont {Ishiguro}},
  \bibinfo {author} {\bibfnamefont {N.}~\bibnamefont {Fukuda}}, \bibinfo
  {author} {\bibfnamefont {Y.}~\bibnamefont {Shimotsuma}}, \ and\ \bibinfo
  {author} {\bibfnamefont {K.}~\bibnamefont {Miura}},\ }\href {\doibase
  10.1364/oe.21.026921} {\bibfield  {journal} {\bibinfo  {journal} {Opt.
  Express}\ }\textbf {\bibinfo {volume} {21}},\ \bibinfo {pages} {26921}
  (\bibinfo {year} {2013})}\BibitemShut {NoStop}%
\bibitem [{\citenamefont {Juodkazis}\ \emph {et~al.}(2006)\citenamefont
  {Juodkazis}, \citenamefont {Nishimura}, \citenamefont {Tanaka}, \citenamefont
  {Misawa}, \citenamefont {Gamaly}, \citenamefont {Luther-Davies},
  \citenamefont {Hallo}, \citenamefont {Nicolai},\ and\ \citenamefont
  {Tikhonchuk}}]{Juodkazis2006}%
  \BibitemOpen
  \bibfield  {author} {\bibinfo {author} {\bibfnamefont {S.}~\bibnamefont
  {Juodkazis}}, \bibinfo {author} {\bibfnamefont {K.}~\bibnamefont
  {Nishimura}}, \bibinfo {author} {\bibfnamefont {S.}~\bibnamefont {Tanaka}},
  \bibinfo {author} {\bibfnamefont {H.}~\bibnamefont {Misawa}}, \bibinfo
  {author} {\bibfnamefont {E.~G.}\ \bibnamefont {Gamaly}}, \bibinfo {author}
  {\bibfnamefont {B.}~\bibnamefont {Luther-Davies}}, \bibinfo {author}
  {\bibfnamefont {L.}~\bibnamefont {Hallo}}, \bibinfo {author} {\bibfnamefont
  {P.}~\bibnamefont {Nicolai}}, \ and\ \bibinfo {author} {\bibfnamefont
  {V.~T.}\ \bibnamefont {Tikhonchuk}},\ }\href@noop {} {\bibfield  {journal}
  {\bibinfo  {journal} {Phys. Rev. Lett.}\ }\textbf {\bibinfo {volume} {96}},\
  \bibinfo {pages} {166101} (\bibinfo {year} {2006})}\BibitemShut {NoStop}%
\bibitem [{\citenamefont {Rapp}\ \emph {et~al.}(2016)\citenamefont {Rapp},
  \citenamefont {Meyer}, \citenamefont {Giust}, \citenamefont {Furfaro},
  \citenamefont {Jacquot}, \citenamefont {Lacourt}, \citenamefont {Dudley},\
  and\ \citenamefont {Courvoisier}}]{Rapp2016}%
  \BibitemOpen
  \bibfield  {author} {\bibinfo {author} {\bibfnamefont {L.}~\bibnamefont
  {Rapp}}, \bibinfo {author} {\bibfnamefont {R.}~\bibnamefont {Meyer}},
  \bibinfo {author} {\bibfnamefont {R.}~\bibnamefont {Giust}}, \bibinfo
  {author} {\bibfnamefont {L.}~\bibnamefont {Furfaro}}, \bibinfo {author}
  {\bibfnamefont {M.}~\bibnamefont {Jacquot}}, \bibinfo {author} {\bibfnamefont
  {P.~A.}\ \bibnamefont {Lacourt}}, \bibinfo {author} {\bibfnamefont {J.~M.}\
  \bibnamefont {Dudley}}, \ and\ \bibinfo {author} {\bibfnamefont
  {F.}~\bibnamefont {Courvoisier}},\ }\href {\doibase 10.1038/srep34286}
  {\bibfield  {journal} {\bibinfo  {journal} {Sci. Rep.}\ }\textbf {\bibinfo
  {volume} {6}},\ \bibinfo {pages} {34286} (\bibinfo {year}
  {2016})}\BibitemShut {NoStop}%
\bibitem [{\citenamefont {G\"{o}tte}\ \emph {et~al.}(2016)\citenamefont
  {G\"{o}tte}, \citenamefont {Winkler}, \citenamefont {Meinl}, \citenamefont
  {Kusserow}, \citenamefont {Zielinski}, \citenamefont {Sarpe}, \citenamefont
  {Senftleben}, \citenamefont {Hillmer},\ and\ \citenamefont
  {Baumert}}]{Gotte2016}%
  \BibitemOpen
  \bibfield  {author} {\bibinfo {author} {\bibfnamefont {N.}~\bibnamefont
  {G\"{o}tte}}, \bibinfo {author} {\bibfnamefont {T.}~\bibnamefont {Winkler}},
  \bibinfo {author} {\bibfnamefont {T.}~\bibnamefont {Meinl}}, \bibinfo
  {author} {\bibfnamefont {T.}~\bibnamefont {Kusserow}}, \bibinfo {author}
  {\bibfnamefont {B.}~\bibnamefont {Zielinski}}, \bibinfo {author}
  {\bibfnamefont {C.}~\bibnamefont {Sarpe}}, \bibinfo {author} {\bibfnamefont
  {A.}~\bibnamefont {Senftleben}}, \bibinfo {author} {\bibfnamefont
  {H.}~\bibnamefont {Hillmer}}, \ and\ \bibinfo {author} {\bibfnamefont
  {T.}~\bibnamefont {Baumert}},\ }\href {\doibase 10.1364/OPTICA.3.000389}
  {\bibfield  {journal} {\bibinfo  {journal} {Optica}\ }\textbf {\bibinfo
  {volume} {3}},\ \bibinfo {pages} {389} (\bibinfo {year} {2016})}\BibitemShut
  {NoStop}%
\bibitem [{\citenamefont {Couairon}\ and\ \citenamefont
  {Mysyrowicz}(2007)}]{Couairon2007}%
  \BibitemOpen
  \bibfield  {author} {\bibinfo {author} {\bibfnamefont {A.}~\bibnamefont
  {Couairon}}\ and\ \bibinfo {author} {\bibfnamefont {A.}~\bibnamefont
  {Mysyrowicz}},\ }\href {\doibase 10.1016/j.physrep.2006.12.005} {\bibfield
  {journal} {\bibinfo  {journal} {Phys. Rep.}\ }\textbf {\bibinfo {volume}
  {441}},\ \bibinfo {pages} {47} (\bibinfo {year} {2007})}\BibitemShut
  {NoStop}%
\bibitem [{\citenamefont {Sudrie}\ \emph {et~al.}(2002)\citenamefont {Sudrie},
  \citenamefont {Couairon}, \citenamefont {Franco}, \citenamefont {Lamouroux},
  \citenamefont {Prade}, \citenamefont {Tzortzakis},\ and\ \citenamefont
  {Mysyrowicz}}]{Sudrie2002}%
  \BibitemOpen
  \bibfield  {author} {\bibinfo {author} {\bibfnamefont {L.}~\bibnamefont
  {Sudrie}}, \bibinfo {author} {\bibfnamefont {A.}~\bibnamefont {Couairon}},
  \bibinfo {author} {\bibfnamefont {M.}~\bibnamefont {Franco}}, \bibinfo
  {author} {\bibfnamefont {B.}~\bibnamefont {Lamouroux}}, \bibinfo {author}
  {\bibfnamefont {B.}~\bibnamefont {Prade}}, \bibinfo {author} {\bibfnamefont
  {S.}~\bibnamefont {Tzortzakis}}, \ and\ \bibinfo {author} {\bibfnamefont
  {A.}~\bibnamefont {Mysyrowicz}},\ }\href {\doibase
  https://doi.org/10.1103/PhysRevLett.89.186601} {\bibfield  {journal}
  {\bibinfo  {journal} {Phys. Rev. Lett.}\ }\textbf {\bibinfo {volume} {89}},\
  \bibinfo {pages} {186601} (\bibinfo {year} {2002})}\BibitemShut {NoStop}%
\bibitem [{\citenamefont {McGloin}\ and\ \citenamefont
  {Dholakia}(2005)}]{McGloin2005}%
  \BibitemOpen
  \bibfield  {author} {\bibinfo {author} {\bibfnamefont {D.}~\bibnamefont
  {McGloin}}\ and\ \bibinfo {author} {\bibfnamefont {K.}~\bibnamefont
  {Dholakia}},\ }\href {\doibase http://dx.doi.org/10.1080/0010751042000275259}
  {\bibfield  {journal} {\bibinfo  {journal} {Contemp. Phys.}\ }\textbf
  {\bibinfo {volume} {46}},\ \bibinfo {pages} {15} (\bibinfo {year}
  {2005})}\BibitemShut {NoStop}%
\bibitem [{\citenamefont {Polesana}\ \emph {et~al.}(2008)\citenamefont
  {Polesana}, \citenamefont {Franco}, \citenamefont {Couairon}, \citenamefont
  {Faccio},\ and\ \citenamefont {Trapani}}]{Polesana2008}%
  \BibitemOpen
  \bibfield  {author} {\bibinfo {author} {\bibfnamefont {P.}~\bibnamefont
  {Polesana}}, \bibinfo {author} {\bibfnamefont {M.}~\bibnamefont {Franco}},
  \bibinfo {author} {\bibfnamefont {A.}~\bibnamefont {Couairon}}, \bibinfo
  {author} {\bibfnamefont {D.}~\bibnamefont {Faccio}}, \ and\ \bibinfo {author}
  {\bibfnamefont {P.~D.}\ \bibnamefont {Trapani}},\ }\href@noop {} {\bibfield
  {journal} {\bibinfo  {journal} {Phys. Rev. A: At., Mol., Opt. Phys.}\
  }\textbf {\bibinfo {volume} {77}} (\bibinfo {year} {2008})}\BibitemShut
  {NoStop}%
\bibitem [{\citenamefont {Bhuyan}\ \emph {et~al.}(2010)\citenamefont {Bhuyan},
  \citenamefont {Courvoisier}, \citenamefont {Lacourt}, \citenamefont
  {Jacquot}, \citenamefont {Salut} \emph {et~al.}}]{Bhuyan2010}%
  \BibitemOpen
  \bibfield  {author} {\bibinfo {author} {\bibfnamefont {M.~K.}\ \bibnamefont
  {Bhuyan}}, \bibinfo {author} {\bibfnamefont {F.}~\bibnamefont {Courvoisier}},
  \bibinfo {author} {\bibfnamefont {P.~A.}\ \bibnamefont {Lacourt}}, \bibinfo
  {author} {\bibfnamefont {M.}~\bibnamefont {Jacquot}}, \bibinfo {author}
  {\bibfnamefont {R.}~\bibnamefont {Salut}},  \emph {et~al.},\ }\href {\doibase
  10.1063/1.3479419} {\bibfield  {journal} {\bibinfo  {journal} {Appl. Phys.
  Lett.}\ }\textbf {\bibinfo {volume} {97}} (\bibinfo {year} {2010}),\
  10.1063/1.3479419}\BibitemShut {NoStop}%
\bibitem [{\citenamefont {Courvoisier}\ \emph {et~al.}(2016)\citenamefont
  {Courvoisier}, \citenamefont {Stoian},\ and\ \citenamefont
  {Couairon}}]{Courvoisier2016}%
  \BibitemOpen
  \bibfield  {author} {\bibinfo {author} {\bibfnamefont {F.}~\bibnamefont
  {Courvoisier}}, \bibinfo {author} {\bibfnamefont {R.}~\bibnamefont {Stoian}},
  \ and\ \bibinfo {author} {\bibfnamefont {A.}~\bibnamefont {Couairon}},\
  }\href {\doibase 10.1016/j.optlastec.2015.11.026} {\bibfield  {journal}
  {\bibinfo  {journal} {Opt. Laser Technol.}\ }\textbf {\bibinfo {volume}
  {80}},\ \bibinfo {pages} {125} (\bibinfo {year} {2016})}\BibitemShut
  {NoStop}%
\bibitem [{\citenamefont {Zhang}(2008)}]{Zhang2008}%
  \BibitemOpen
  \bibfield  {author} {\bibinfo {author} {\bibfnamefont {Y.}~\bibnamefont
  {Zhang}},\ }\href {\doibase 10.1007/s00340-007-2830-4} {\bibfield  {journal}
  {\bibinfo  {journal} {Applied Physics B: Lasers and Optics}\ }\textbf
  {\bibinfo {volume} {90}},\ \bibinfo {pages} {93} (\bibinfo {year}
  {2008})}\BibitemShut {NoStop}%
\bibitem [{\citenamefont {Goodman}(1996)}]{Goodman1996}%
  \BibitemOpen
  \bibfield  {author} {\bibinfo {author} {\bibfnamefont {J.~W.}\ \bibnamefont
  {Goodman}},\ }\href@noop {} {\emph {\bibinfo {title} {Introduction to Fourier
  optics}}}\ (\bibinfo  {publisher} {McGraw-Hill},\ \bibinfo {year}
  {1996})\BibitemShut {NoStop}%
\bibitem [{\citenamefont {Collins}\ and\ \citenamefont
  {O'Connor}(2015)}]{Collins2015}%
  \BibitemOpen
  \bibfield  {author} {\bibinfo {author} {\bibfnamefont {A.~R.}\ \bibnamefont
  {Collins}}\ and\ \bibinfo {author} {\bibfnamefont {G.~M.}\ \bibnamefont
  {O'Connor}},\ }\href {\doibase 10.1364/OL.40.004811} {\bibfield  {journal}
  {\bibinfo  {journal} {Opt. Lett.}\ }\textbf {\bibinfo {volume} {40}},\
  \bibinfo {pages} {4811} (\bibinfo {year} {2015})}\BibitemShut {NoStop}%
\bibitem [{\citenamefont {Young}\ and\ \citenamefont
  {Budynas}(2002)}]{Young2002}%
  \BibitemOpen
  \bibfield  {author} {\bibinfo {author} {\bibfnamefont {W.~C.}\ \bibnamefont
  {Young}}\ and\ \bibinfo {author} {\bibfnamefont {R.~G.}\ \bibnamefont
  {Budynas}},\ }\href@noop {} {\emph {\bibinfo {title} {Roark's formulas for
  stress and strain}}},\ \bibinfo {number} {7th edition}\ (\bibinfo
  {publisher} {McGraw-Hill New York},\ \bibinfo {year} {2002})\BibitemShut
  {NoStop}%
\bibitem [{\citenamefont {Nattermann}\ \emph {et~al.}(1998)\citenamefont
  {Nattermann}, \citenamefont {Neuroth},\ and\ \citenamefont
  {Scheller}}]{Nattermann1998}%
  \BibitemOpen
  \bibfield  {author} {\bibinfo {author} {\bibfnamefont {K.}~\bibnamefont
  {Nattermann}}, \bibinfo {author} {\bibfnamefont {N.}~\bibnamefont {Neuroth}},
  \ and\ \bibinfo {author} {\bibfnamefont {R.~J.}\ \bibnamefont {Scheller}},\
  }\enquote {\bibinfo {title} {The properties of optical glass},}\ \ (\bibinfo
  {publisher} {Springer-Verlag Berlin Heidelberg},\ \bibinfo {year} {1998})\
  Chap.\ \bibinfo {chapter} {Mechanical Properties}, pp.\ \bibinfo {pages} {pp.
  179--200}\BibitemShut {NoStop}%
\end{thebibliography}%

\end{document}